\documentclass[aps,prl,twocolumn,showpacs,groupedaddress]{revtex4}
\usepackage{graphicx}
\usepackage{dcolumn}
\usepackage{bm}

\newcommand{\eins}{\mbox{$1 \hspace{-1.0mm} {\bf l}$}}

\begin{document}


\title{Decoherence-Free Quantum Information Processing with\\
Four-Photon Entangled States}
\author{
Mohamed Bourennane$^{1,2}$,
Manfred Eibl$^{1,2}$,
Sascha Gaertner$^{1,2}$,
Christian Kurtsiefer$^{2}$,\\
Ad\'{a}n Cabello$^{3}$, and
Harald Weinfurter$^{1,2}$}
\affiliation{
$^{1}$Max-Planck-Institut f\"{u}r Quantenoptik,
D-85748 Garching, Germany\\
$^{2}$Sektion Physik, Ludwig-Maximilians-Universit\"{a}t,
D-80797 M\"{u}nchen, Germany\\
$^{3}$Departamento de F\'{\i}sica Aplicada II, Universidad de Sevilla,
E-41012 Sevilla, Spain}


\begin{abstract}
Decoherence-free states protect quantum information from
collective noise, the predominant cause of decoherence in current
implementations of quantum communication and computation. Here we
demonstrate that spontaneous parametric down-conversion can be
used to generate four-photon states which enable the encoding of
one qubit in a decoherence-free subspace. The immunity against
noise is verified by quantum state tomography of the encoded
qubit. We show that particular states of the encoded qubit can be
distinguished by local measurements on the four photons only.
\end{abstract}



\maketitle

\date{\today}

\newpage


Quantum information processing enables secure classical
communication, powerful quantum communication schemes, and speedup
in computation~\cite{QC00}. These methods rely on the preparation,
manipulation, and detection of the superposition of quantum
states. Superpositions, however, are very fragile and easily
destroyed by the decoherence processes due to unwanted coupling
with the environment~\cite{Zurek91}. Such uncontrollable
influences cause noise in the communication or errors in the
outcome of a computation, and thus reduce the advantages of
quantum information methods.

Several strategies have been devised to cope with decoherence,
each of them appropriate for a specific type of coupling with the
environment. For instance, if the interaction with the environment
is weak enough such that qubits are affected only with a very low
probability, a good strategy would be to add redundancy when
encoding the quantum information in order to detect and correct
the errors by active quantum error correction methods~\cite{QECC}.

If the qubit-environment interaction, no matter how strong,
exhibits some symmetry, then there exist quantum states which are
invariant under this interaction. These states are called
decoherence-free (DF) states, and allow protection of quantum
information \cite{PSE96,ZR97a,Lidar,KBLW01}. A particularly
relevant symmetry arises when the environment couples with the
qubits without distinguishing between them, resulting in the
so-called collective noise. This situation occurs, for instance,
when the spatial (temporal) separation between the carriers of the
qubits is small relative to the correlation length (time) of the
environment. Typical examples arise in ion-trap or nuclear
magnetic resonance (NMR) experiments, which are susceptible to
fluctuations of magnetic or electrostatic fields, but also in
quantum communication, e.g., when the qubits carried by polarized
photons are successively sent via the same optical fiber and
therefore experience the same birefringence.


Experimental efforts investigating features of DF systems so far
have been limited to two-qubit systems only. For two qubits,
however, the singlet state is the only DF state and thus, it is
not sufficient to fully protect an {\em arbitrary} logical qubit
against collective noise. The experiments so far have demonstrated
the features of the single DF state or the immunity against
restrictive types of noise \cite{df2}. For three qubits there is
no DF state immune to collective noise. However, quantum
information can be preserved in an abstract subsystem known as a
noiseless subsystem, which was demonstrated in NMR experiments
\cite{df3}.

In this Letter we report on the production of various
decoherence-free four-photon polarization-entangled states using
spontaneous parametric down conversion (SPDC). The immunity of
these states against collective noise is experimentally verified
by showing their invariance when passing the four photons through
a noisy environment simulated by birefringent media. Moreover, we
show that one can both distinguish two orthogonal four-photon DF
states and perform state tomography by local polarization
measurements only. We are thus able to demonstrate that quantum
information can be reliably extracted from qubits communicated
through a noisy environment, and also between parties who do not
even share a common reference frame.

For the construction of DF states formed by $N$ qubits, we note
that for collective noise, all $N$ qubits undergo the same
(unknown) unitary transformation $U$. States are decoherence-free
if they are invariant under such a $N$-lateral unitary
transformation, i.e., $ U^{\otimes N} |\psi\rangle= |\psi\rangle$,
where $U^{\otimes N} = U\otimes...\otimes U$ denotes the tensor
product of $N$ unitary transformations $U$ \cite{ZR97a}. The
amount of quantum information that a given DF subspace is able to
protect is determined by the number $N$ of qubits
used~\cite{ZR97a}. For $N=2$ qubits there is only one DF state,
the singlet state, $|\psi^-\rangle_{ab} = (1/\sqrt{2})
(|01\rangle-|10\rangle)_{ab}$, where $|01\rangle_{ab}=|0\rangle_a
\otimes |1\rangle_b$. The smallest useful DF subspace is spanned
by two four-qubit DF states. We can choose one of them as the
tensor product of two singlet states,
\begin{equation}
|\Phi_0\rangle_{abcd}=|\psi^-\rangle_{ab}\otimes
|\psi^-\rangle_{cd}.
\label{phi0}
\end{equation}
The DF state orthogonal to $|\Phi_0\rangle_{abcd}$ is given by
\begin{eqnarray}
|\Phi_1\rangle_{abcd} & = & {1 \over 2 \sqrt{3}}
(2|0011\rangle-|0101\rangle-|0110\rangle -|1001\rangle
\nonumber \\
& &
-|1010\rangle+2|1100\rangle )_{abcd},
\label{phi1}
\end{eqnarray}
and was first introduced by Kempe {\it et al.}~\cite{KBLW01}.

The DF subspace spanned by the states $|\Phi_0\rangle$ and
$|\Phi_1\rangle$ allows now to encode a qubit $|\Phi\rangle=c_0
|0\rangle+c_1 |1\rangle$ (where $c_0$ and $c_1$ are complex
numbers) as the superposition state
$|\Phi_L\rangle=c_0|\Phi_0\rangle+c_1 |\Phi_1\rangle$, which is
immune against collective noise for any $c_0$ and $c_1$. It is
also important to be able to read the logical qubit
$|\Phi_L\rangle$. This is usually carried out by projecting
$|\Phi_L\rangle$ onto the basis states of the DF subspace
requiring non trivial quantum gates. However, $|\Phi_0\rangle$ and
$|\Phi_1\rangle$ can be distinguished using only local
measurements: It suffices to project the first two qubits onto the
computational basis ($|0\rangle$ and $|1\rangle$) and the other
two onto the Hadamard rotated basis [$|\bar{0}\rangle=(|0\rangle +
|1\rangle)/\sqrt{2}$ and $|\bar{1}\rangle=(|0\rangle -
|1\rangle)/\sqrt{2}$]. Expressed in these bases, the DF states
read as
\begin{eqnarray}
|\Phi_0\rangle & = &
(|01\bar{1}\bar{0}\rangle-|01\bar{0}\bar{1}\rangle
+|10\bar{0}\bar{1}\rangle-|10\bar{1}\bar{0}\rangle)/2,
\label{zzxx0} \\
|\Phi_1\rangle & = &
(|00\bar{0}\bar{0}\rangle-|00\bar{0}\bar{1}\rangle
-|00\bar{1}\bar{0}\rangle+|00\bar{1}\bar{1}\rangle
-|01\bar{0}\bar{0}\rangle
\nonumber \\ & &
+|01\bar{1}\bar{1}\rangle-|10\bar{0}\bar{0}\rangle
+|10\bar{1}\bar{1}+|11\bar{0}\bar{0}\rangle
+|11\bar{0}\bar{1}\rangle
\rangle \nonumber \\ & &
+|11\bar{1}\bar{0}\rangle+|11\bar{1}\bar{1}\rangle)/2 \sqrt{3}.
\label{zzxx1}
\end{eqnarray}

These states have no common terms and are therefore easily
distinguishable using the outcomes of local measurements on the
four qubits.

The invariance of the encoded quantum information is demonstrated
best by comparing the density matrix $\rho_L =
|\Phi_L\rangle\langle \Phi_L|$ of the logical qubit before
($\rho_{{\rm in}}$) and after ($\rho_{{\rm out}}$) the interaction
with the environment. In order to evaluate the density matrix
$\rho$ of an encoded qubit, one needs to measure 3 four-qubit
observables $\Sigma_z$, $\Sigma_x$, and $\Sigma_y$. A well-suited
choice is $\Sigma_z =
\sigma_z\otimes\sigma_z\otimes\sigma_x\otimes\sigma_x$, $\Sigma_x
=\sigma_z\otimes\sigma_x\otimes\sigma_z\otimes\sigma_x$, and
$\Sigma_y =\sigma_y\otimes\sigma_x\otimes\sigma_z\otimes\eins$,
because they can be determined again by local measurements on the
four photons. Here, $\{\sigma_x,\sigma_y,\sigma_z\}$ denote the
Pauli matrices, and $\eins$ is the identity. The results of these
measurements allow us to perform the tomographic reconstruction of
the density matrix $\rho$, since its elements can be expressed as
$\rho_{11} = (3\langle\Sigma_z\rangle + 1)/4$, ${\rm
Re}(\rho_{12}) = \sqrt{3}(2\langle\Sigma_x\rangle +
\langle\Sigma_z\rangle - 1)/4$, and ${\rm Im}(\rho_{12}) =
\sqrt{3}\langle\Sigma_y\rangle/2$, where $\langle \Sigma_i
\rangle={\rm Tr}\left({\rho\Sigma_i}\right)$ describes the
expectation value of $\Sigma_i$.


\begin{figure}[b]
\centering
\includegraphics[{width=8.0cm}]{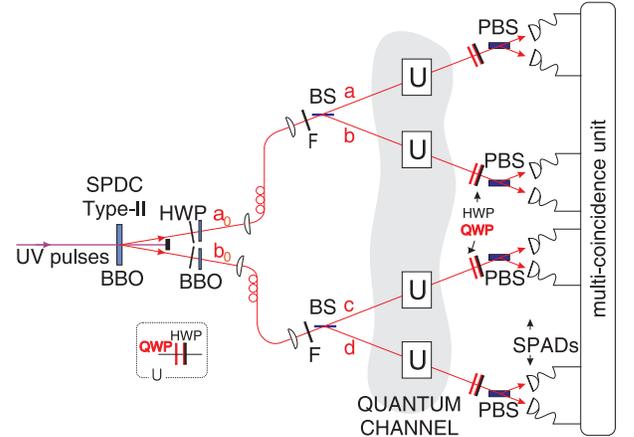}
\vspace{0.5cm} \caption{Experimental setup to show the invariance
of four-photon entangled states under collective noise. The
photons are emitted from spontaneous parametric down-conversion in
a BBO crystal followed by birefringence compensation into two
spatial modes $a_0$ and $b_0$. They are distributed into the four
modes $a$, $b$, $c$, $d$ by 50:50 beam splitters (BS) behind
interference filters (F). The noisy quantum channel causing the
unitary transformation $U^{\otimes4}= U \otimes U \otimes U
\otimes U$ is simulated by equal combinations of quarter- (QWP)
and half-wave plates (HWP). Additional waveplates and polarizing
beamsplitters (PBS) are employed for the polarization analysis of
the four photons. For the registration of the decoherence-free
states, events are selected where one photon was detected by
single-photon avalanche diodes (SPAD) in each of the four modes. }
\end{figure}


In our experiment the physical qubits are polarized photons, where
the computational basis corresponds to horizontal and vertical
linear polarization, ``0'' $\equiv H$ and ``1''$\equiv V$. The
four-photon polarization-entangled state $|\Phi_0 \rangle$ can be
obtained from two synchronized (first order) SPDC sources for
photon pairs in the singlet state $|\psi^- \rangle$. More
practically, for the measurements shown in Figs.~2 and ~3,
$|\Phi_0 \rangle$ was generated by using the product state of two
polarization-entangled photon pairs created from two consecutive
pump pulses and swapping the modes $b \leftrightarrow c$. The
four-photon polarization-entangled state $|\Phi_1 \rangle$ was
observed using the second order SPDC process \cite{WZ01,EGBKZW03}.

We used the UV pulses of a frequency-doubled mode-locked
Ti:sapphire laser (pulse length 130\,fs) to pump a 2\,mm thick BBO
(barium betaborate) crystal at a wavelength of 390\,nm and a
repetition rate of 82\,MHz with an average power of 750\,mW (see
Fig.~1). The pump beam was focused to a waist of 100\,$\mu$m
inside the crystal. The degenerate down-conversion emission into
the two characteristic type-II crossing directions~\cite{KWIAT95}
was coupled into single mode optical fibers and passed through
narrowband interference filters ($\Delta\lambda=3$\,nm) to exactly
define the spatial and spectral emission modes. To observe
$|\Phi_0 \rangle$ and $|\Phi_1 \rangle$, two
polarization-independent 50:50 beam splitters were used to split
the four photons into the four modes $a$, $b$, $c$, and $d$. Next,
the photons were sent through the quantum channel, where the noisy
environment was simulated by a combination of birefringent
quarter- (QWP) and half- (HWP) wave plates in each arm. The
polarization analysis was performed using further waveplates and
polarizing beam splitters followed by silicon avalanche single
photon detectors. Only events with one photon detected in each of
the four arms have been selected.

Figure 2 shows the~16 possible fourfold coincidences for
polarization analysis of one photon in each of the four outputs of
the beam splitters exhibiting the characteristic statistics of the
states $|\Phi_0\rangle$ (A) and $|\Phi_1\rangle$ (B). As a measure
of the quality of the state preparation we use the quantum bit
error rate (QBER), which is defined as the ratio of false events
over total events or, in terms of the four-photon visibility
$\textit{V}$ \cite{EGBKZW03}, as ${\rm QBER}=(1-\textit{V})/2$.
For the data shown, we obtain ${\rm QBER}=3.91\% \pm 0.44 \%$ (A)
and ${\rm QBER} =4.30\% \pm 0.25\%$ (B). The ratio of total events
observed upon encoding $|\Phi_1\rangle$ and $|\Phi_0\rangle$ is
expected to be 3 for otherwise similar pumping conditions. Within
the variation in the observed four-photon rate, this ratio is also
reflected in the experiment.


\begin{figure}[t]
\centering
\includegraphics[width=8.0cm]{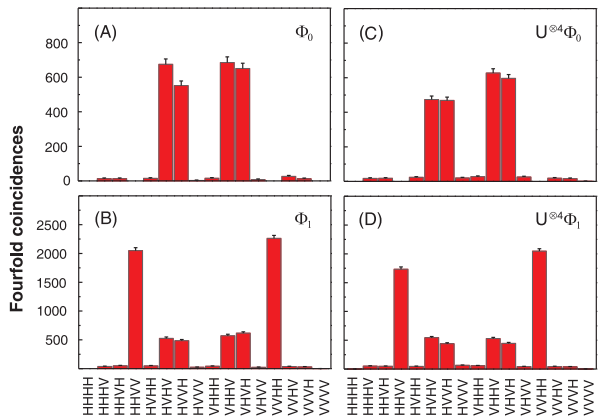}
\caption{Total counts of all the 16 possible fourfold detection
events in $4$\,h measurement time for the four-photon states
$|\Phi_{0}\rangle$~(A) and $|\Phi_{1}\rangle$~(B). Labels H and V
indicate the horizontal and vertical polarization measurement
outcomes for the four photons. (C) and (D) show the results in the
presence of collective noise, i.e., under the same unitary
transformation $U$. Here and in the following, $U$ was arbitrarily
chosen and set with a HWP at an angle of $59^{\circ}$ and a QWP at
an angle of $13.5^{\circ}$. }
\end{figure}


To demonstrate the invariance of the four-photon states $|\Phi_0
\rangle$ and $|\Phi_1 \rangle$ under collective decoherence, i.e.,
under phase and bit flip errors caused by a birefringent quantum
channel, we have arbitrarily chosen the unitary transformation $U
= 0.012i\eins -0.332\sigma_z - 0.707\sigma_y + 0.624\sigma_x$.
We implement this unitary transformation here by the addition of
a HWP set at an angle of $59^{\circ}$ and a QWP set at an
angle of $13.5^{\circ}$.

Figures~2(C) and 2(D) show that the distribution of detection
events is not changed for the states $|\Phi_0 \rangle$ and
$|\Phi_1 \rangle$ under the unitary transformation $U^{\otimes4}$
(i.e., when the four photons are subject to collective noise).
Here, we obtain similar error rates, with ${\rm QBER}=7.11\% \pm
0.50\%$ (C), and ${\rm QBER} = 6.41\% \pm 0.28\%$ (D) for $|\Phi_0
\rangle$ and $|\Phi_1 \rangle$, respectively. From these
measurements one can deduce the diagonal elements of the
four-photon density matrices. Obviously, no additional elements
are populated under the action of the collective noise indicating
that the states $|\Phi_0 \rangle$ and $|\Phi_1 \rangle$ do not
leave the DF subspace.


\begin{figure}[t]
\centering
\includegraphics[width=8.0cm]{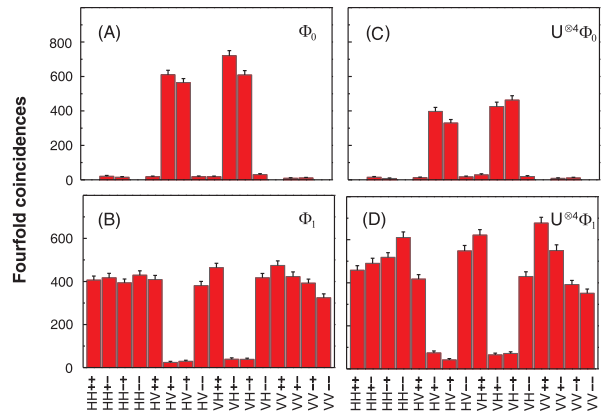}
\caption{Fourfold coincidences distinguishing between the states
$|\Phi_0\rangle$ and $|\Phi_1\rangle$ by local polarization
measurements on the four photons (counts in $4$\,h). The photon
polarization is analyzed in the $\{H,V\}$ and $\{+,-\}$ (i.e.,
$\pm45^{\circ}$ basis) for the photons in paths ($a$, $b$) and
($c$, $d$), respectively. As before, (A) and (B) show the results
of the analysis of the states $|\Phi_0\rangle$ and
$|\Phi_1\rangle$ without noise, whereas (C) and (D) are the
results in the presence of the collective noise $U^{\otimes4}$. }
\end{figure}


To distinguish between the states $|\Phi_0 \rangle$ and $|\Phi_1
\rangle$ by local measurements, we have projected the photon
polarizations in paths $a$ and $b$ on the $\{H,V\}$ basis, and in
paths $c$ and $d$ on the $\{+,-\}$, i.e., $\pm45^{\circ}$
polarization basis. Figures~3(A) and 3(B) show the fourfold
coincidence counts corresponding to a detection of $|\Phi_0
\rangle$ and $|\Phi_1 \rangle$, respectively, for a noiseless
environment. For $|\Phi_1 \rangle$, we clearly observe that the
false fourfold coincidence counts, i.e., the terms $HV+-$, $HV-+$,
$VH+-$, and $VH-+$, are negligible compared to the other terms,
and vice versa for $|\Phi_0 \rangle$. We observe an error rate of
${\rm QBER}= 5.23\% \pm 0.46\%$ for $|\Phi_0 \rangle$ (A) and
${\rm QBER}= 2.56\% \pm 0.22\%$ for $|\Phi_1 \rangle$ (B).
Analyzing the DF states in presence of the collective noise
$U^{\otimes4}$ shows that one is still able to distinguish
reliably the two DF states, now with ${\rm QBER}= 6.82\% \pm 0.75
\%$ (C) and ${\rm QBER}= 3.99\% \pm 0.26\%$ (D), respectively.

In order to encode any arbitrary logical qubit, one could use two
sources, one for $|\Phi_0 \rangle$ and the other for $|\Phi_1
\rangle$, and coherently overlap the generated photons. Yet, the
technical requirements go beyond a first proof of principle. For a
demonstration of the invariance of a logical qubit encoded in DF
states, we prepared the state $|\Psi_L \rangle = (\sqrt{3}|\Phi_0
\rangle - |\Phi_1 \rangle)/2$ and performed quantum state
tomography of the encoded qubit before and after passage through a
noisy quantum channel. We choose this state as it can be also
obtained from the set-up shown in Fig.~1 by swapping modes $b
\leftrightarrow c$. Figure~4 shows the elements of the density
matrices $\rho_{{\rm in}}$ (A) and $\rho_{{\rm out}}$ (B) in the
$\{|\Phi_0\rangle,|\Phi_1\rangle\}$ basis of the logical qubit.
The imaginary parts of the density matrices obtained are
negligible. These results show that the diagonal elements before
and after the interaction are in good agreement, and that the
relative phase between basis states $ |\Phi_0 \rangle$ and
$|\Phi_1 \rangle$ is conserved. The quality of preparation of
$\rho_{\rm in}$ into the desired state $\rho_L =
|\Phi_L\rangle\langle \Phi_L|$ is characterized by the fidelity
$F_{\rho_{\rm in},\rho_L} = {\rm Tr}[(\sqrt{\rho_{L}}\rho_{\rm
in}\sqrt{\rho_{L}})^{1/2}]$, where we obtain an experimental value
of $F_{\rho_{\rm in}, \rho_L} = 0.989 \pm 0.038 $. After exposing
the state to the collective noise $U^{\otimes4}$ we obtain an
overlap between the initial and the outgoing state $F_{\rho_{\rm
in}, \rho_{\rm out}} = 0.996 \pm 0.076$, showing that the quantum
information encoded in a DF subspace is preserved.

In addition, we want to point out that besides protecting against
collective noise, the DF states are useful also for the
communication of quantum information between two observers who do
not share a common reference frame~\cite{BRS03}. In such a
scenario, any realignment of the receiver's reference frame
corresponds to the application of the same unitary transformation
to each of the qubits which were sent. Yet, such an operation does
not affect $|\Phi_0\rangle$, $|\Phi_1\rangle$, or any
superposition thereof. Therefore, it is irrelevant whether or not
the receiver's reference frame is aligned with the sender's
reference frame in order to read the quantum information encoded
in the DF states. Let us assume that the misalignment between
sender and receiver is just given by the unitary transformation
$U$. In this case the results shown in Fig.~3 and Fig.~4 clearly
demonstrate that the receiver obtains the correct quantum
information even if his reference frame does not coincide with the
one of the sender.


\begin{figure}[t]
\centering
\includegraphics[width=8.0cm]{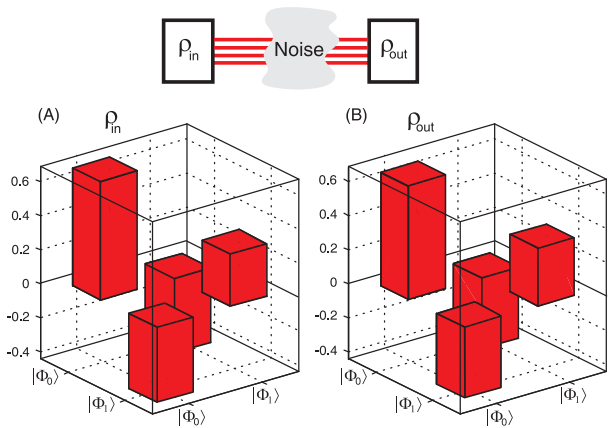}
\caption{Propagation of the logical qubit $|\Psi_L \rangle =
(\sqrt{3}|\Phi_0 \rangle - |\Phi_1 \rangle)/2$: (A) and (B) show
the experimentally obtained density matrices before ($\rho_{\rm
{in}}$) and after ($\rho_{\rm {out}}$) passage through a noisy
quantum channel. The encoding in a DF subspace protected the
transmission, leading to a fidelity of $F_{\rho_{\rm in},
\rho_{\rm out}} = 0.9958 \pm 0.0759$ in the presence of noise
(overall measurement time 12\,h).} \label{fig4}
\end{figure}


To summarize, we experimentally demonstrated that SPDC can
directly produce four-photon entangled states required to encode
quantum information in decoherence-free subspaces and to protect
it against collective noise. The quantum information encoded in
the DF subspace is accessible by local measurements of the four
photons, without two-qubit quantum logic gates being necessary,
and thus realizable with state-of-the-art technology. This is
relevant for possible applications of quantum communication
\cite{qkd-df}. We have performed a tomographic reconstruction of
the density matrix of a logical qubit encoded in the DF subspace
showing its immunity against a noisy environment. Our measurements
also show that DF states permit the communication of quantum
information even if the sender and the receiver do not share a
reference frame. This is of great importance for future
experiments studying quantum nonlocal effects between distant
observers \cite{BRS03,Adan03}.


This work was supported by the DFG (We 2451/1-2), the EU-Project
RamboQ (IST-2001-38864), the Spanish MCYT Project BFM2002-02815,
and the ESF.


\end{document}